# TD-BPQBC: A 1.8μW 5.5mm$^3$ ADC-less Neural Implant SoC utilizing 13.2pJ/Sample Time-domain Bi-phasic Quasi-static Brain Communication

Baibhab Chatterjee, K Gaurav Kumar, Shulan Xiao, Gourab Barik, Krishna Jayant and Shreyas Sen
Purdue University, West Lafayette, Indiana 47907, USA. Email: {bchatte, gauravk, xiao208, gbarik, kjayant, shreyas}@purdue.edu

***Abstract*—Untethered miniaturized wireless neural sensor nodes with data transmission and energy harvesting capabilities call for circuit and system-level innovations to enable ultra-low energy deep implants for brain-machine interfaces. *Realizing that the energy and size constraints of a neural implant motivate highly asymmetric system design* (a small, low-power sensor and transmitter at the implant, with a relatively higher power receiver at a body-worn hub), we present Time-Domain Bi-Phasic Quasi-static Brain Communication (TD-BPQBC), offloading the burden of analog to digital conversion (ADC) and digital signal processing (DSP) to the receiver. The input analog signal is converted to time-domain pulse-width modulated (PWM) waveforms, and transmitted using the recently developed BPQBC method for reducing communication power in implants. The overall SoC consumes only 1.8μW power while sensing and communicating at 800kSps. The transmitter energy efficiency is only 1.1pJ/b, which is >30X better than the state-of-the-art, enabling a fully-electrical, energy-harvested, and connected in-brain sensor/stimulator node.**

## I. Introduction

### *A. Background and Related Work*

In recent years, several miniaturized wireless neural sensor nodes [1]-[9] have been demonstrated for data communication and powering in brain-machine interfaces (BMIc). Most of them use RF, Inductive, Optical (OP), Ultrasound (US) and Magneto-Electric (ME) techniques, which either suffer from significant tissue absorption (RF/Inductive methods), or transduction losses (OP, US and ME techniques). The high amount of tissue absorption in traditional radiative RF requires large transmission power for system-level operation. For example, [9] uses a transmit power of 0.5W, which exceeds ICNIRP safety guidelines [10] by ~10X. US [1] or OP [3], [8] telemetry are safer, at the cost of significant loss (up to 110 dB loss as shown in [1]) due to scattering and skull absorption, requiring a sub-cranial interrogator which needs to be surgically placed, and reduces end-to-end efficiency. Magneto-Electric (ME) [7] methods exhibit extremely low body-absorption, but has large transduction loss (0.1mT magnetic field requirement in [7] which is equivalent to ~300kV/m electric field for iso-energy-density), lowering the end-to-end efficiency. Capacitive body-coupled powering and communication techniques [11]-[12] have been shown to work well with wearable devices, but the same techniques are not proven conclusively for implants.

Bi-Phasic Quasistatic Brain Communication (BPQBC) [4],[6] has recently been shown as a promising alternative for powering and data transmission in such implants which utilizes fully-electrical quasi-static signaling to transmit signals from a brain implant to an external hub placed on the skull/skin, with low end-to-end channel loss (<60dB for 50mm channel length, avoiding any challenge due to transduction loss) and ≈52pJ/b effective energy efficiency due to digital data compression. However, [4] assumes that the communication is narrowband and an ADC precedes the BPQBC transmitter (TX), generating the digital bits from neural signals as shown in Fig. 1 (bottom left). A compressive Sensing (CS) DSP core helps in reducing the number of bits to be transmitted, thereby reducing the communication burden. Both the ADC and the DSP are power-hungry, and thus the TX needs to be heavily duty cycled (0.1% as shown in [4], without even considering the ADC power). This duty cycle limitation results from the requirement of operating the implants with harvested energy, which is usually in the range of a few μW when powered from wearable sources, within safety limits for human-centric applications [10].

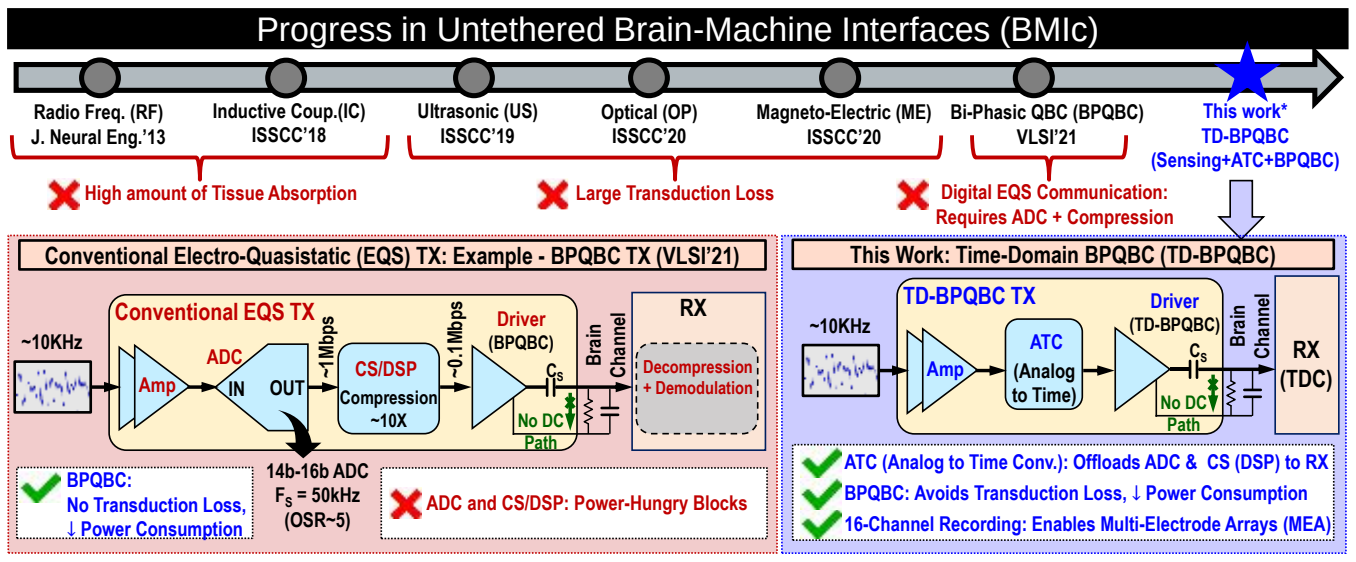

Fig. 1. Progress in Untethered BMIc, leading to Time-domain BPQBC (TD-BPQBC): data transmission from a brain implant to a body-worn hub, with analog information being encoded in pulse-width of a PWM signal.

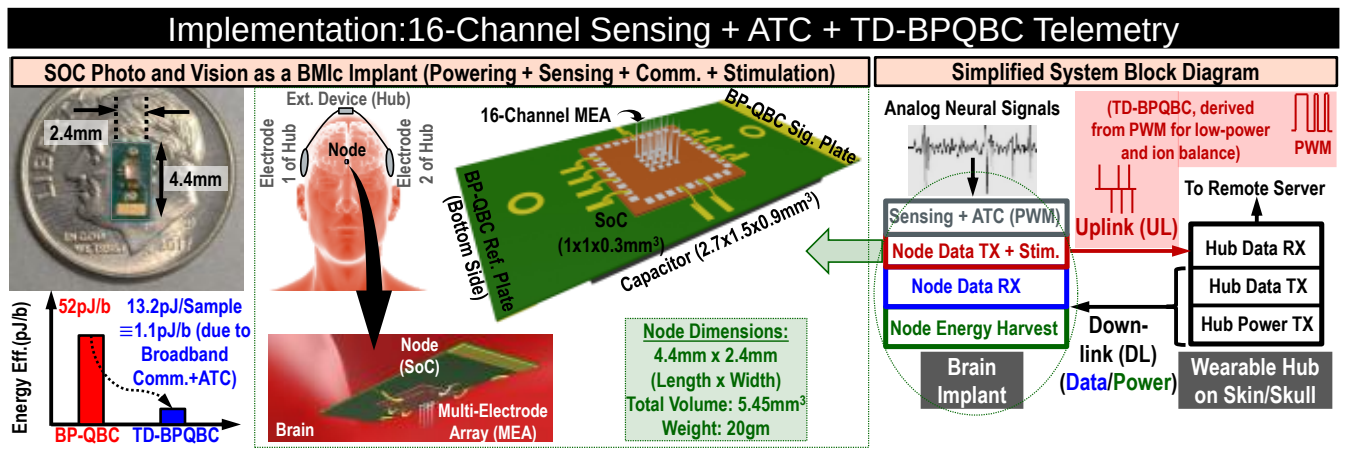

Fig. 2. Details of the node implementation and performance summary.

### *B. Our Contribution*

Realizing that the energy and size constraints of a neural implant motivate highly asymmetric system design, we present, for the first time, Time-Domain BPQBC (TD-BPQBC), which offloads the burden of ADC and DSP to the body-worn receiver (RX), while transmitting the input analog signal by converting it to time-domain using an analog to time converter (ATC), as shown in Fig. 1 (bottom right). The ATC converts an analog neural signal to a pulse-width modulated (PWM) wave, which is conducive toward digital-friendly BPQBC techniques due to its 2-level signaling, leading to ultra-low-power. In this work, we implemented a 5.5 mm$^3$ sensor and a 1.1pJ/b TX at the implant that sends the TD-

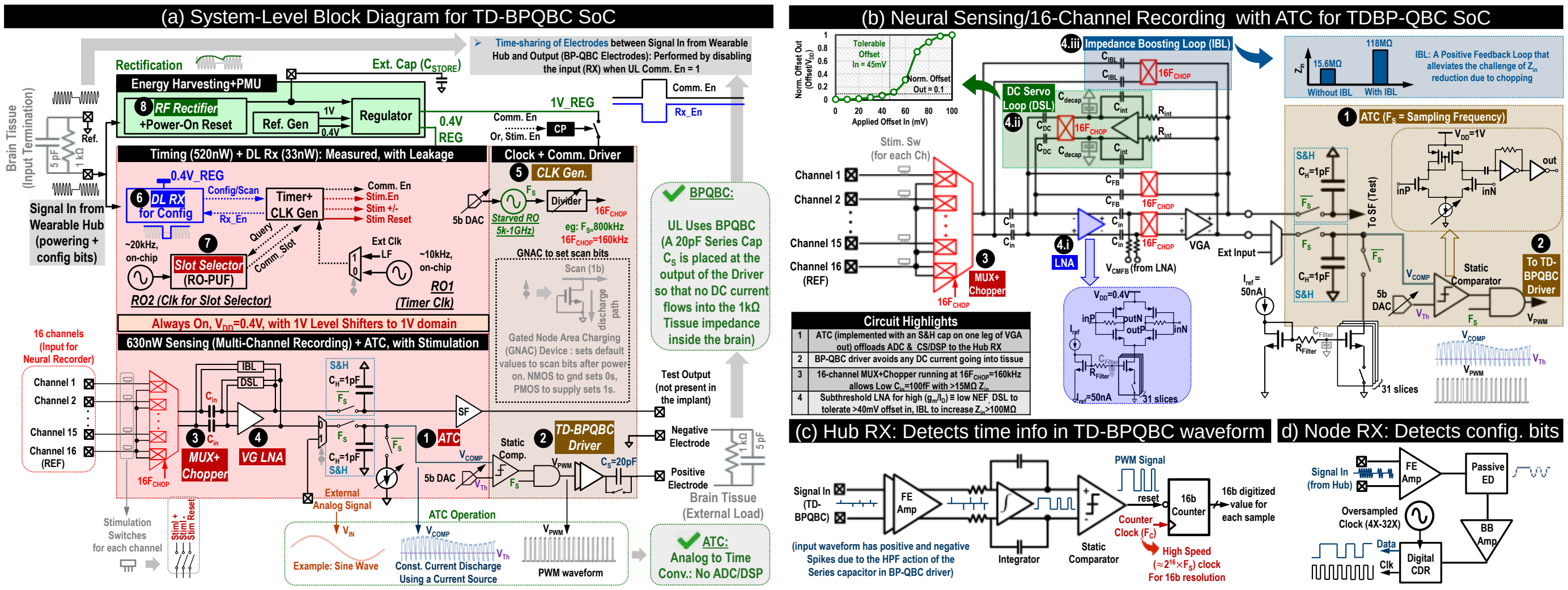

Fig. 3. (a) System Diagram of TD-BPQBC SoC, with 1) a 120nW ATC (analog to time conversion), 2) a 9.8µW TD-BPQBC driver, 3) combined 16MUX+Chopper front end, 4) a variable gain (40-60dB) LNA, along with auxiliary circuits: 5) variable frequency clock generator, 6) downlink RX, 7) Communication slot selector preventing collision, 8) an RF rectifier for energy harvesting, with power-on reset; (b) Circuit Diagram and Implementation Highlights of the 16-channel Neural Recording Front-end with ATC and BPQBC; (c) Hub UL RX and (d) BMIc node DL RX design.

BPQBC waveform to a relatively higher power receiver at a body-worn hub, forming an uplink (UL). The downlink (DL) path is from the wearable hub to the node, which is used to transfer power to the node and for configuring the implant.

Hence, the *key contributions* of this work are as follows: (1) This is the *first implementation of time-domain fully-electrical brain communication*, showing that energy efficiencies of up to 1.1pJ/b is possible (>30X improvement over literature); (2) Analysis of *offloading the burden of ADC and DSP to the UL RX* is presented, in terms of the trade-off in achievable signal to distortion ratio (SNDR) vs. sampling frequency at the ATC and the resolution of the time to digital converter (TDC) at the UL RX, showing that a high sampling frequency (~800kHz) as well as a moderate to high TDC bit-resolution (>12 bits) is required for < 2dB degradation in SNDR as compared to the input; (3) Development of a 5.5mm$^3$ Neural SoC for simultaneous powering, multi-channel sensing, stimulation and communication. Together, these key contributions lead to the *first fully electrical, energy-harvested, and connected in-brain sensor/stimulator node, with orders of magnitude lower powering requirement* compared to other brain communication modalities that often require field transduction (conversion between electrical energy and optical/ultrasound/magnetic energies, resulting in higher end-to-end system loss).

## II. Operating Principles and Circuit Design

As shown in Fig. 2, the implementation involves a brain implant/node with 1) 16-channel sensing, 2) ATC, 3) an UL TD-BPQBC data TX, 4) bi-phasic current stimulation (500nA-1mA), 5) a DL TD-BPQBC data RX and 6) energy harvesting. The node sends the TD-BPQBC signals through the brain tissue, and communicates with a hub (placed on the skin/ skull) containing 1) an UL data RX and 2) a DL power and data TX. The use of duty cycling with a variable communication slot selector ensures that two nodes are unlikely to communicate to the hub at the same time (similar to [4]), and hence broadband communication can be implemented. This improves the 52pJ/b energy efficiency in BPQBC to 13.2pJ/Sample in TD-BPQBC, while the ATC improves it further to 1.1pJ/b (>45x improvement vs. [4]) by detecting 12b/sample in the UL hub RX.

The 65nm TD-BPQBC SoC implementation is shown in Fig. 3(a), featuring 1) an 120nW ATC for converting the analog information to time, 2) a 9.8µW TD-BPQBC driver to reduce power consumption and maintain ion balance, 3) a combined MUX+Chopping front end for 16 input channels, 4) a variable gain (40-60dB) low-noise amplifier (LNA), along with auxiliary circuits for clocking, DL RX, communication slot selection and powering. These features are described in detail in Fig. 3(b). The 16-channel neural recording front-end, interfacing with 16 on-chip electrodes (please see die photo as shown in Fig. 4), utilizes a combined MUX+Chopping stage with 16X higher clock frequency ($16F_{CHOP}$) than that is required for a single channel ($F_{CHOP}$). The use of a higher chopping frequency through the use of combined MUX+Chopper ensures that the succeeding LNA stage can be designed with small (W/L) transistors and small AC-coupling capacitors ($C_{in}$≈100fF), which keeps the input impedance ($Z_{in}$) of the chopping front-end to acceptable values (>15MΩ) even without any impedance boosting loop (IBL). Similar to [13], a separate positive feedback IBL increases the $Z_{in}$ to > 118MΩ by providing the required charging currents during chopping. A DC servo loop (DSL) helps in tolerating up to 45mV of DC electrode offset. The LNA is designed with subthreshold transistors to 1) reduce the power consumption and 2) design at a better ($g_m/I_D$) point for a better noise efficiency factor (NEF). The 120nW ATC is implemented by 1) passing the output of the amplifier through a sample and hold (S&H) with a

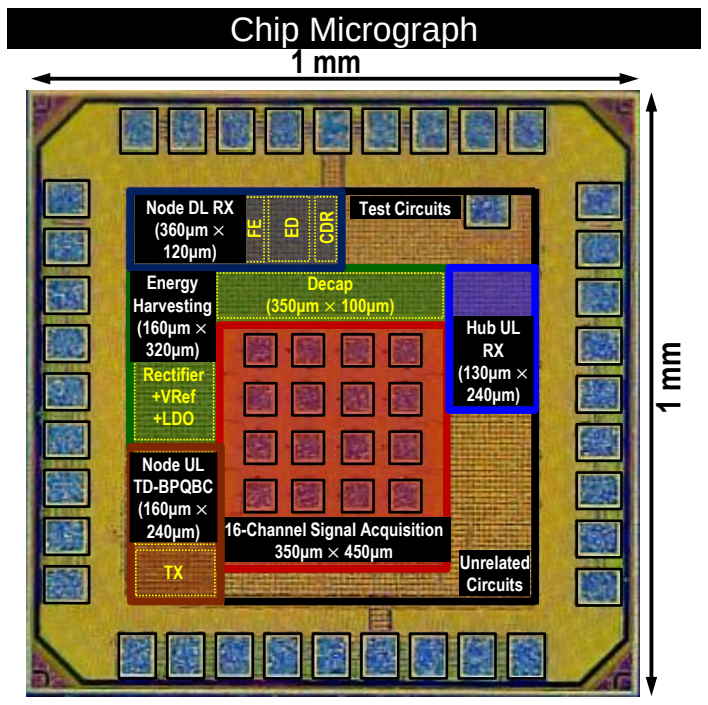

Technology Specs and Design Summary

| | | |
|---|---|---|
| Test Chip | Process | 65nm 1P9M CMOS LP |
| | Package | Gold Wire-bond with Glob-Top Encapsulation, Polyimide Flex PCB |
| UL TD-BPQBC TX + 16-Channel Sensing | Supply | 0.8V-1V |
| | Active Area | 0.2mm$^2$ (0.16mm$^2$ for 16-channel sensing, 0.04mm$^2$ for the UL TX) |
| | Load | 1kΩ Res. \|\| 5pF Cap |
| | Performance | 13.2pJ/Sample at 800kSps |
| | IO used | 21 + 11 IOs for Scan/Test outputs |
| DL RX (UL RX numbers are in parenthesis) | Supply | 0.4V (1V) |
| | Active Area | 0.0432mm$^2$ (0.0312mm$^2$) |
| | i/p Termination | 1kΩ Res. \|\| 5pF Cap (20pF cap) |
| | Performance | 33pJ/b@1kbps (688pJ/b@9.6Mbps) |
| | IO used | 3 (4) + 7 IOs for Scan/Test outputs |
| Energy Harvest | Active Area | 0.0862mm$^2$ |
| | IO used | 2 + 3 IOs for Scan/Test outputs |

Fig. 4. Die micrograph of the SoC and Design Summary.

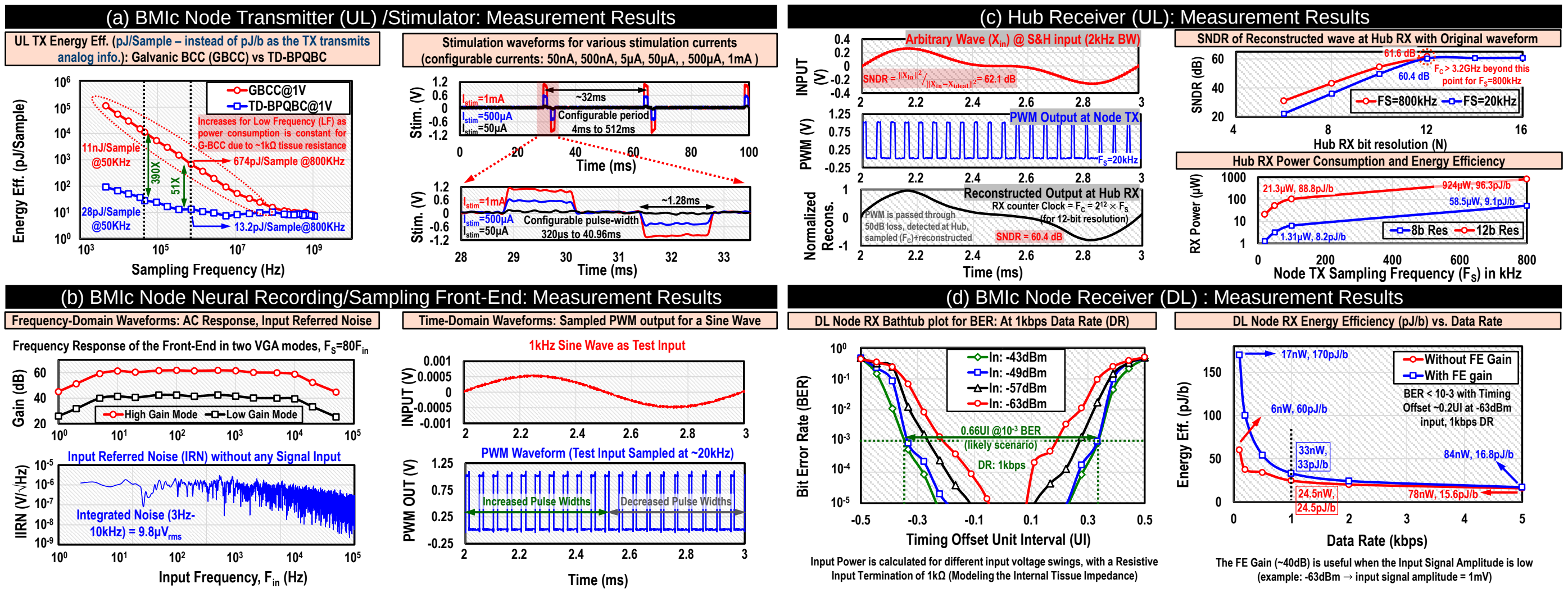

Fig. 5. (a) Measurement for TD-BPQBC UL TX and stimulator: 51X better energy efficiency than GBCC @800kSps; (b) Measurement for front-end, with 1) 40-60dB gain, 2) 9.8μVrms noise, 3) time-domain waveforms; (c) Measurement for UL Hub RX: reconstructed output with $<$ 2dB degradation in SNDR w.r.t. original wave if Hub RX bit res. $\geq$ 12-bit; (d) DL Node RX performance in terms of BER and Energy Efficiency, showing 33pJ/b efficiency at 1kbps

sampling frequency of $F_S \approx 800$kHz, 2) discharging the S&H capacitor ($C_H$) during the hold mode through a constant current discharge, and 3) compare the linear discharging waveforms with a constant voltage using a static comparator, thereby generating a PWM waveform, which is provided to the TD-BPQBC driver. By sharing the hold mode with the discharge cycle, this method requires only one capacitor for ATC ($C_H \approx 1$pF). Similar to BPQBC [4], the TD-BPQBC driver at the UL TX blocks DC current paths into the brain tissue by use of a 20pF series capacitor ($C_S$), which is large enough to ensure that ≈80% of the driver AC voltage is available at the ≈5pF load as presented by the brain tissue (obtained through FEM simulations in HFSS using the NEVA EM Human v2.2 model), but is still small enough (100μm × 100μm) for on-chip implementations. Fig. 3(c)-(d) show the UL hub RX and DL node RX topologies. The UL RX uses an integrator to compensate for the high-pass filtered (differentiated) TD-BPQBC waveform due to $C_S$, and utilizes a 16b counter with a high-speed clock ($F_C = 2^N \times F_S$) to get N-bit resolution for each sample at the UL RX. Please note that the ATC's current and capacitances will determine the full scale at the UL RX, while the comparator in the UL TX may introduce a fixed offset for the PWM signal, which is taken care of using a 5-bit digital offset compensation circuit. The DL RX detects configuration bits from the DL hub TX to program the node.

## III. Measurement Results

Fig. 5(a) shows the measured results for the UL TX and the stimulator (3 different current configs), while Fig. 5(b) presents the results for the recording/sampling front-end. Compared to traditional Galvanic Body Channel Communication (GBCC), TD-BPQBC shows improved energy efficiencies (51X better at 800kHz and 390X better at 50kHz) as the DC current path to the tissue is blocked by $C_S$. The recording front end has 40-60dB programmable gain and an integrated input referred noise of 9.8μ$V_{rms}$. Fig. 5(c)-(d) show the performance of the UL hub RX and DL node RX. The UL hub RX receives reduced-amplitude PWM signals (with ~50dB loss representing the body channel) with analog information residing on the edge locations. Since the implant is ADC-less, the overall performance of the ATC system is measured by a calculating the SNDR (which is a ratio of the L2 norms of the signal to noise and distortion), both at the input of the ATC and at the reconstructed analog waveform at the RX. The moderate aliasing distortion present in the reconstructed signal in Fig. 5(c) (left-hand-side) can be removed by using a low-pass filter with sharp roll-off. For $F_S$ = 20kHz, the SNDR degrades from 62.1dB at the input to 60.4dB at the reconstructed waveform at the RX, when 12-bit resolution is used at the UL RX (N=12, $F_C = 2^{12} \times F_S$). However, the SNDR at the reconstructed waveform depends on both $F_S$ and RX bit-resolution (N). A higher $F_S$ results in a better SNDR because of a higher oversampling ratio. Similarly, a higher N will result in a better SNDR, till the point it is saturated with the input SNDR (~62.1dB). As shown in Fig. 5(c) (right-hand-side), by using a higher $F_S$ = 800kHz, reconstructed SNDR becomes 61.6dB for N = 12. Please note that with $F_S$ = 800kHz, SNDR beyond N $>$ 12 was not measured as $F_C >$ 3.2GHz at this point. At $F_S$ = 800kHz, the energy efficiency for the UL RX is 9.1pJ/b for N=8 and 96.3pJ/b for N=12. The degradation in energy efficiency from N=8 to N=12 is expected, as we only achieve a linearly increasing bit-resolution at the cost of an exponentially increasing $F_C$ at the UL RX, which results in an overall degradation in pJ/b. However, in terms of the UL TX, this asymmetric TD-BPQBC link enables the 1.8μW Neural Node SoC with 5% duty cycling. Out of the 1.8μW, only 0.49μW is consumed in the duty-cycled TX. Without any duty cycling, the node will consume 11.1μW.

With 5% duty cycling, and N=12 at the UL RX, this results in an UL TX energy efficiency of 1.1pJ/b, which is a $>$ 45x improvement over [4]. As shown in Fig. 5(d), the bit error rate (BER) for the DL RX at 1kbps data rate (DR) with 31-bit-PRBS shows a timing margin of >0.66UI for a BER of $10^{-3}$ at -43dBm input (10m$V_{pp}$, calculated with 1kΩ $R_{in}$ || 5pF $C_{in}$, i.e. the effective termination in the brain tissue [4]). The energy consumption of the DL RX is 33pJ/b with a front-end (FE) on-chip amplifier, and 24.5pJ/b without the FE amplifier at 1kbps.

The Full-System Block Diagram for the TD-BPQBC SoC is shown in Fig. 6(a). The 5.5$mm^3$ node is placed on the brain of a C57BL/6J Mouse, adhering to the overseeing Animal Care and Use Committee guidelines. Two RX electrodes (signal+ref) are placed on the surface of the head. The Node communicates the signal through the Brain Tissue, and is received at the hub where an N-bit counter/TDC converts the PWM waveform to digital bits.

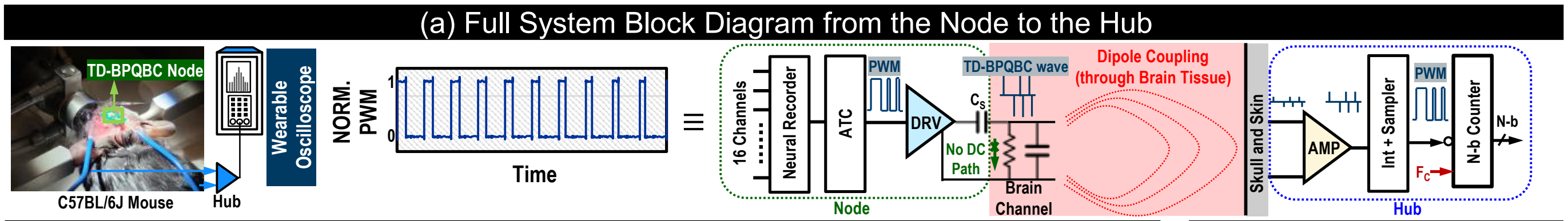

(b) Comparison with state-of-the-art BMIc Communication Systems

| | | This Work (TDBP-QBC, aided by ATC) | Previous Literature on BMIc Communication | | | | |
|---|---|---|---|---|---|---|---|
| | | | Nat. Electron'21 Lee [5] | VLSI'21 Chatterjee [4] | ISSCC'20, Lim [3] | SSCL'20, Yuk [2] | ISSCC'19, Ghanbari [1] |
| System | Process Node (nm) | 65 | 65 | 65 | 180 | 110 | 65 |
| System | Link Architecture | BP-QBC (ATC based**) | RF | BP-QBC (Narrowband) | Optical (Narrowband) | Capacitive BCC* (Broadband**) | Ultra-sonic (Narrowband) |
| System | Chip Size (mm²) | 1 x 1 | 0.65 x 0.65 | 1 x 1 | 0.19 x 0.17 | 1 X 0.5 | 0.5 x 0.5 |
| System | Implant Volume (mm³) | 5.45 | 0.1 (only the IC) | 5.54 | N/A | N/A | 0.8 |
| System | External Components | 1 cap + Electrode | 1 Relay Coil, | 1 cap + Electrode | 1 LED + 1 PV Diode + Electrode | 1 Battery + LDOs + Electrode | 1 Piezo + Electrode |
| System | Sub-Cranial Interrogator | Not required | Not required | Not required | Yes | Not required | Yes |
| System | End-to-End loss | ~60dB (@55mm) | - | ~60dB (@55mm) | >80dB | ~40dB (@30mm) | >80dB |
| System | Standby SoC Power when not communicating (μW) | 1.3 ( sensing+ATC+leakage in comm. module) | - | 0.66 (leakage in comm. module) | N/A | N/A | 13 (leakage in comm. module) |
| System | Average SoC Power (μW) | 1.8§ | - | 1.18§§ | > 0.74 | > 566.1 | 41.8 |
| UL | UL (Node to Hub) Communication | PWM (ATC), BP-QBC | BPSK | OOK, BP-QBC | Sym. Inter-val Mod. | 3-level Polar RZ, Capacitive BCC | AM back-scatter |
| UL | Comm. Collision Resolution | Yes: RO-PUF based | Yes | Yes | Yes | No | Yes |
| UL | UL Node TX Power (Active) | 0.49μW@800kSps§ | - | 0.52μW@10Mbps§§ | 0.74μW | 566.1μW** | 28.8μW |
| UL | Encoded Data Rate (DR) | $DR_{effective}$= 480kbps (12b/ sample, 5% duty cycling) | - | $DR_{effective}$=10kbps (0.1% duty cycling) | 100bps | 16.7Mbps (Broadband) | N/A |
| UL | UL Node TX Energy Efficiency | 1.1pJ/b@800kSps (12b/sample) | - | 52pJ/b@10Mbps | >7.4nJ/ LED firing | 33.9pJ/b** @16.7Mbps | N/A |
| DL | DL Node RX Power | 33nW (1kbps) | - | 31nW (1kbps) | N/A | N/A | No DL |
| DL | DL Node Clock Recovery | Yes | - | Yes | Yes | Yes | No |

* Utilizes capacitive coupling through the parasitic return path of external components, which would lead to lower signal reeption in a truly implantable scenario
§: With 5% duty cycling. Digitization happens at Hub RX §§: With CS and 0.1%duty cycling ** No carrier involved

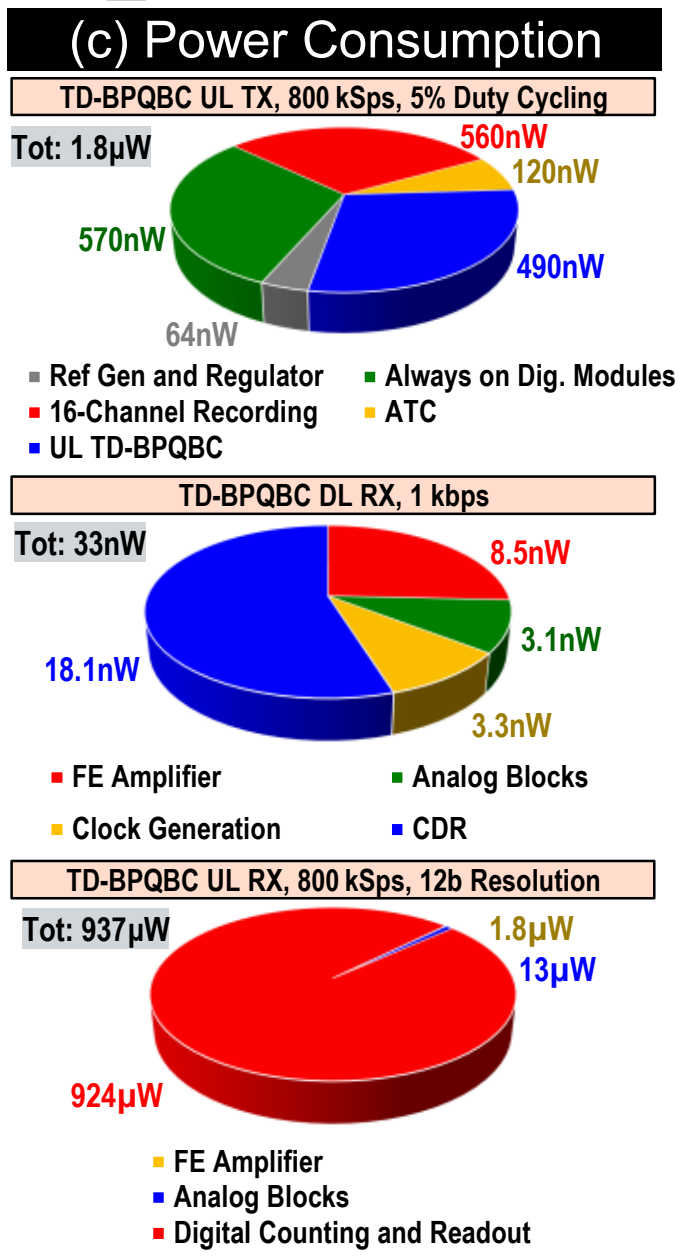

Fig. 6. (a) Vision of the full system block diagram from the Node to the Hub; (b) state-of-the-art BMIc telemetry SoC comparison, showing TD-BPQBC's >30X better energy efficient UL TX as compared to [2],[4]; (c) Power Consumption Summary for UL TX (in the Node), DL RX (in the Node) and UL RX (in the Hub).

## IV. Comparison with State-of-the-Art

Fig. 6(b) compares the TD-BPQBC SoC performance with state-of-the-art BMIc telemetry systems, exhibiting the best energy efficiency (30X improvement vs. [2], and >45X vs. [4]), while preserving the benefits of BPQBC, namely, 1) max. channel length (up to 55mm) and 2) no transduction loss due to fully-electrical quasi-static signaling with information embedded in the time-domain. Although optical and ultrasonic techniques can result in small implant sizes, the low end-to-end system loss of the TD-BPQBC method would allow room for further miniaturization.

## V. Conclusions and Future Work

The implemented SoC is the first *ADC-less TD-BPQBC link* for simultaneous powering, 16-channel sensing, data communication and stimulation. The combination of ATC with TD-BPQBC results in ~1.1pJ/b energy efficiencies at the UL TX, while a UL RX with TDC bit resolution > 12 can reconstruct the analog signal with < 2dB degradation in the SNDR, when a sampling frequency of 800kHz is used in the ATC of the UL TX. The trade-offs involving power, RX bit-resolution, sampling frequency and SNDR are analyzed. The presented SoC is the first fully electrical, energy-harvested, and connected in-brain sensor/stimulator node, with orders of magnitude lower powering requirement compared to other brain communication modalities that often require field transduction. This will potentially motivate future ADC-less asymmetric network architectures for energy-efficient brain implants. As future work, body-coupled galvanic and bi-phasic modalities of power transfer, and further miniaturization for such mm-scale implants will be explored in further detail.

## Acknowledgment

This work was supported by Quasistatics, Inc. (Grant #40003567)